\documentclass[9pt,twocolumn,twoside,dvipsnames]{opticajnl}
\journal{opticajournal} 

\setboolean{shortarticle}{true}

\usepackage{svg}
\usepackage{lineno}
\usepackage{amsmath}
\usepackage{pdfpages}

\definecolor{mydarkblue}{rgb}{0., 0., 0.35}
\usepackage{xparse} 
\usepackage[normalem]{ulem} 
\usepackage{soul} 

\dates{To be published in Optics Letters, Vol.~50, Issue~21, pp.~6445--6448 (2025)}
\doi{Optics Letters link: \url{https://doi.org/10.1364/OL.566273}}

\NewDocumentCommand{\revision}{m}{#1}
\NewDocumentCommand{\replace}{mm}{#2} 

\newcommand{\fdop}{f_{\text{dop}}}

\newcommand{\ihet}{i_{\text{het}}}

\title{Bayesian approach for spatial super-resolution of heterodyne wind lidars}

\author[1,*]{Theo Martin}
\author[2]{Laurent M. Mugnier}
\author[1]{Matthieu Valla}
\author[3]{Pierre Etienne Allain}
\author[1]{David Tomline Michel}

\affil[1]{DOTA, ONERA, Université Paris-Saclay, 91120, Palaiseau, France}
\affil[2]{DOTA, ONERA, Université Paris-Saclay, 92320, Châtillon, France}
\affil[3]{Vaisala France, Tech Park, 6A Rue René Razel, 91400 Saclay, France}

\affil[*]{theo.martin@onera.fr}

\begin{abstract}
Wind speed measurements using heterodyne lidars are limited in spatial resolution because of the current signal processing methods. This limit is equal to $c\tau$ ($c$ the speed of light and $\tau$ the laser pulse duration) corresponding to the length of atmosphere contributing to the wind speed measurement at one distance. To go beyond this limit, we use an inverse problem approach based on a model of the spectrogram (concatenation of periodograms of each range) and prior distributions on our unknowns: backscattering amplitude and wind speed at each range. We apply our inversion method on simulated and experimental spectrograms demonstrating a gain in resolution of a factor from 2 to 2.5 depending on the signal to noise ratio.
\end{abstract}

\setboolean{displaycopyright}{false} 

\begin{document}

\maketitle
\section{Introduction}
 
\begin{figure}[h!tb]
\begin{center}\leavevmode
\includegraphics[width=0.8\linewidth]{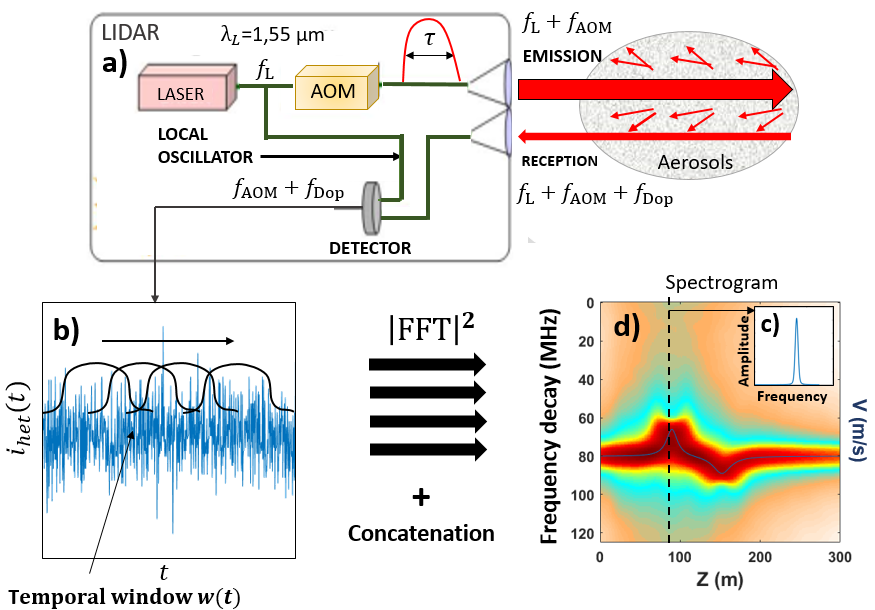}
\vspace*{-0.5\baselineskip}
  \caption{Schematic of wind heterodyne lidar. a) Lidar architecture b) Example of heterodyne signal c) Periodogram from spectrogram d) Spectrogram of a vortex. FFT: Fast Fourier Transform}
  \label{LidarSchema}
  \end{center}
\end{figure}

The study of the wind field is important in areas such as aviation to measure wind gusts over airports~\cite{BeaAirport}~\cite{TheseMatthieu}, wind energy to optimize the efficiency of wind turbines~\cite{WindEnergy}, and meteorology~\cite{AppliMeteo}. An instrument for measuring wind speed is the pulsed coherent Doppler wind lidar. Its advantages are spatially resolved measurements along the line of sight and long-range measurements over 10 km. The whole lidar process is represented in Figure~\ref{LidarSchema}. 
To obtain the wind speed at one point at a distance $z$ in the atmosphere, the lidar emits a laser pulse of duration $\tau$ at a typical wavelength of $\lambda = 1.55\,$µm. An acousto-optic modulator (AOM) shifts the laser frequency by $f_{\text{AOM}}$. The pulse is backscattered by the aerosols and shifted by a frequency $\fdop = -2v/\lambda$, where $v$ is the projected wind speed on the line of sight. The frequency shift is the consequence of the Doppler effect induced by the relative aerosols’ speed. The backscattered light returns to the lidar after a time $2z/c$, with $c$ the speed of light, and is mixed on the detector with a part of the laser source called a local oscillator to perform the heterodyne detection. Then, by taking a subset of the heterodyne signal on a duration $\tau$ using a temporal window, the periodogram is computed, and the peak frequency of the spectrum, corresponding to the Doppler frequency, can be estimated. 

Due to the duration $\tau$ of the laser pulse, the signal contributing to one point in the heterodyne signal comes from a region of atmosphere of length $c\tau/2$. Moreover, using the temporal window to analyze the heterodyne signal over a duration $\tau$, the region of atmosphere contributing is increased by $c\tau/2$. Finally, the spatial resolution of pulsed wind lidars is limited by $c\tau$. Shortening the laser pulse to improve spatial resolution introduces design and engineering challenges. A shorter pulse has a spectral broadening that degrades the spectral resolution. Also, it has less energy because the peak power is limited in the fibers by Brillouin scattering~\cite{smith1972optical}. So, the range is limited, and the signal-to-noise ratio is decreased. Emitting pulses more frequently to improve the mean power limits the lidar range because of the ambiguity distance, and using a more powerful laser has a cost, bigger size, and energy consumption.

To estimate the frequency of the peak, the simplest estimator is the centroid or barycenter estimator~\cite{harris2006lidar}. A more precise method is the Gaussian estimator~\cite{GaussianFit}, also known as parabolic fit or polynomial fit, which uses the Gaussian shape of the spectrum to fit a second-order polynomial on the logarithm of the spectrum. Other more efficient estimators are the Maximum Likelihood Estimator (MLE)~\cite{MLE} and the Even-Order Derivative Sharpening Technique (EDPST)~\cite{EDPST} which is useful for very short pulses. The main limit of these methods is the independent processing of the spectra, which has the inherent spatial resolution limit of $c\tau$. To remove this limit some signal processing methods have been proposed, such as a 2D Wiener deconvolution~\cite{Wiener2DLidar} of the spectrogram (concatenation of the spectrum along the line of sight) using the impulse response of the lidar and a convolution model~\cite{Gretsi}, but without an experimental demonstration of the spatial resolution gain. Another is a joint time-frequency analysis~\cite{JTFA} that shows a gain in spatial resolution \revision{which is not estimated and only compared to an undersampled reference}. Moreover, the wind speed estimation is noisy even under good signal-to-noise ratio (SNR) conditions. In addition, the last two methods require manual setup of a regularization parameter.

Other methods have been developed to improve spatial resolution, but involve a modification of the hardware, such as pulse modulation~\cite{GolayCoding}~\cite{SubMeterPRCPM}, or differential spectrum between two pairs of pulses~\cite{PSK}. These methods require additional arbitrary wave generator and electro-optic modulators, significantly increasing the complexity of lidar hardware.

In this paper, we show for the first time an unsupervised method to improve the spatial resolution of a wind lidar based on a MAP inversion of the spectrogram. We study the spatial resolution gain depending on the SNR both in simulations and experimentally.

\section{Spectrogram inversion}

This section presents our spectrogram inversion method. In the first part, we show that the spectrogram can be expressed as a convolution. Then we present the criterion for our Maximum A Posteriori (MAP) method with physically sound regularizations on the unknowns.

\subsection{Direct model of spectrogram}

The raw data available in the lidar output is the heterodyne signal $\ihet$. The heterodyne signal has two terms: the first is the heterodyne beat from the difference between the local oscillator and the backscattered signal from the atmosphere, which has a frequency $f_{\text{AOM}} + \fdop$. The second is the noise due to the shot noise from the local oscillator on the detector. We pick a subset of the signal using a window $w$ at a certain time and compute the periodogram  (squared modulus of the Fast Fourier transform (FFT)). Because the periodogram of a single laser shot is very noisy, we average thousands of laser shots to improve the signal-to-noise ratio and therefore the frequency peak estimation. We concatenate the periodogram computed at different distances along a line of sight to form the spectrogram $S_{\text{exp}}$. This computation reads:
\begin{equation}
    S_{\text{exp}}(z_i,f_l)=\biggl< \left| \text{FFT}_{t_n}\left[ w\left( t_n - \frac{2z}{c}\right) \ihet(t_n,k)\right] \right|^2 (f_l) \biggr>_k ,
\end{equation}
where $k$ is the k-th laser shot. We discretize the distances $z$ into a regularly spaced set ${z_i}$, and ${f_l}$ are the frequency samples because the spectrum is computed with a finite number of points.
Assuming the cross terms between noise and signal of the periodogram are negligible after an average over thousands of laser shots and using the "feuilleté" model of the atmosphere~\cite{salamitou1995simulation}, the model spectrogram $S$ can be written as a discrete version of the convolutive model we have derived in~\cite{Gretsi}:
\begin{equation}
    S(z_i,f_l \ ;\boldsymbol{A} , \boldsymbol{\nu})=
    \sum_{i',l'} PSF(z_i\!-\!z_{i'},f_l\!-\!f_{l'})
    A(z_{i'})\text{sinc}\left(f_{l'}\!-\!\nu(z_{i'})
    \right) \, ,
    \label{Model}
\end{equation}

where $PSF(z_i,f_l)$ (Point Spread Function) is the spectrogram of a still point target located at $z=0$, $\boldsymbol{A}$ and $\boldsymbol{\nu}$ are the vectors of unknowns of the backscattering amplitudes and of the frequency shifts.
The $\text{sinc}$ is the discrete version of the Dirac $\delta({f=\nu(z)})$.

\subsection{Inversion method}

Our goal is to estimate the backscattering amplitudes $\boldsymbol{A}$ and the frequency shifts $\boldsymbol{\nu}$. As we want to add physical knowledge to our solution, we will use a MAP method \revision{(see, \emph{e.g.},} \protect\cite[Chap.~3,]{idier2013bayesian} \revision{for an introduction to Bayesian estimation and MAP)} \replace{defined as follows}{and we define the solution as}:
\begin{align}
    [\hat{\boldsymbol{A}},\hat{\boldsymbol{\nu}}]&=\text{arg}\ \underset{\boldsymbol{A},\boldsymbol{\nu}}{\text{max}} \ [ p(S_{\text{exp}}|(\boldsymbol{A},\boldsymbol{\nu})) \times p(\boldsymbol{A}) \times  p(\boldsymbol{\nu}) ] \ , \\
    [\hat{\boldsymbol{A}},\hat{\boldsymbol{\nu}}]&=  \text{arg}\ \underset{\boldsymbol{A},\boldsymbol{\nu}}{\text{min}} \ [K(\boldsymbol{A},\boldsymbol{\nu})+\mathcal{R}_A(\boldsymbol{A})+\mathcal{R}_{\nu}(\boldsymbol{\nu})] \ ,
\end{align}
where $p(S_{\text{exp}}|(\boldsymbol{A},\boldsymbol{\nu}))$ is the likelihood of the data $S_{\text{exp}}$, $ p(\boldsymbol{A})$, $p(\boldsymbol{\nu})$ are the \textit{a priori} probability distributions for the unknwons $\boldsymbol{A}$ and $\boldsymbol{\nu}$ respectively, \revision{$K(\boldsymbol{A},\boldsymbol{\nu})=-\text{ln}(p(S_{\text{exp}}|(\boldsymbol{A},\boldsymbol{\nu})))$ is the opposite of the log-likelihood of the data, $\mathcal{R}_A(\boldsymbol{A})=-\text{ln}(p(\boldsymbol{A}))$ and $\mathcal{R}_\nu(\boldsymbol{\nu})=-\text{ln}(p(\boldsymbol{\nu}))$ are the regularization terms for $\boldsymbol{A}$ and $\boldsymbol{\nu}$ respectively.}

We assume a centered stationary Gaussian prior probability distribution for $\boldsymbol{\nu}$. Therefore, the regularization term $\mathcal{R}_\nu$ of $\boldsymbol{\nu}$ can be written \cite{Conan-a-98} as: 
\begin{equation}
    \mathcal{R}_\nu(\boldsymbol{\nu})=-\text{ln}(p(\boldsymbol{\nu}))=\frac{1}{2}\sum_{\gamma}{\frac{|\Tilde{\boldsymbol{\nu}}(\gamma)|^2}{P_{\boldsymbol{\nu}}(\gamma)}} \ ,
\end{equation}
where $\tilde{.}$ denotes the Fourier transform and $P_{\boldsymbol{\nu}}$ is the \emph{a priori} Power Spectral Density of the frequency shifts.

The regularization on $\boldsymbol{\nu}$ is necessary to avoid an uncontrolled noise amplification, especially in the high spatial frequencies (see Supplement Section~1 for more details). To this end, the Kolmogorov law is a reasonable prior to express the \emph{a priori} decrease of energy with increasing spatial frequencies: $P_\nu(\gamma) \propto \gamma^{-5/3}$ where $\gamma$ is the spatial frequency.
By approximating $-5/3 \approx -2$ and using the Parseval theorem, we can write the regularization $\mathcal{R}_\nu$ in the direct domain as the \revision{energy of the} derivative of $\boldsymbol{\nu}$ :
\begin{equation}
    \mathcal{R}_\nu(\boldsymbol{\nu})= \frac{\mu}{2} \vert \vert \nabla \boldsymbol{\nu} \vert \vert ^2 \ ,
\end{equation}
where $\nabla$ is the discrete gradient operator and $\mu$ is a constant, called the frequency shifts $\boldsymbol{\nu}$ regularization hyperparameter.

For the backscattering amplitudes $\boldsymbol{A}$ we assume their profile over the range is very smooth, and we impose this smoothness through the following second-order quadratic regularization:
\begin{equation}
    \mathcal{R}_A (\boldsymbol{A}) = \frac{\eta}{2} \vert \vert \Delta \boldsymbol{A} \vert \vert ^2 = -\text{ln}(p(\boldsymbol{A})) \ ,
\end{equation}
where $\Delta$ is the discrete Laplace operator defined by the kernel [1 -2 1], and $\eta$ is the backscattering amplitudes $\boldsymbol{A}$ regularization hyperparameter. 
The regularization hyperparameters $\eta$ and $\mu$ can be computed without supervision as proposed in other contexts~\cite{paul2013}~\cite{jonquiere2025zonal}. In practice, this computation only requires the estimates of $\| \Delta\boldsymbol{A}\|^2$ and $\| \nabla\boldsymbol{\nu}\|^2$ obtained with the Gaussian estimator.

With the assumption that the data spectrogram is perturbed by an inhomogeneous centered Gaussian noise, the log-likelihood is the sum of weighted square differences between the data spectrogram and the model spectrogram \cite[10.1.4]{idier2013bayesian}. The weights are the inverses of the local noise variances $\sigma^2(z,f)$. This noise can be shown to follow a $\chi^2$-distribution of $N$ (number of laser shots averaged) degrees of freedom and of mean $S(\boldsymbol{A}_{\text{true}} , \boldsymbol{\nu}_{\text{true}})$, the model spectrogram computed from the true unknowns. Since $N$ is generally of the order of several thousand, this distribution can be approximated by a Gaussian distribution. The MAP estimator is the couple $(\boldsymbol{A},\boldsymbol{\nu})$ that minimizes the following criterion $J$:
\begin{align}
    J(\boldsymbol{A},\boldsymbol{\nu})= &\sum_i\sum_l |S_{\text{exp}}(z_i,f_l)-S(z_i,f_l \ ;\boldsymbol{A} , \boldsymbol{\nu})|^2/(2\sigma^2(z_i,f_l)) \nonumber \\
    & +  \ \mathcal{R}_A (\boldsymbol{A}) + \ \mathcal{R}_\nu(\boldsymbol{\nu}) ,
    \label{Criterion}
\end{align}

\subsection{Computation of optimal backscattering amplitudes}

Using the spectrogram model (\ref{Model}) and the fact that the discrete convolution is a sum, we can write the model spectrogram as:

\begin{equation}
   S(z_i,f_l\ ; \boldsymbol{A},\boldsymbol{\nu})=\sum_k PSF(z_i-z_k,f_l-\nu_k)A(k)
\end{equation}
For a given frequency $f_l$ we can therefore express the vector $S(\cdot,f_l|\boldsymbol{\nu},\boldsymbol{A})$ as the product between a matrix $M_l$ and $\boldsymbol{A}$:

\begin{equation}
    S(\cdot,f_l|\boldsymbol{A},\boldsymbol{\nu})=M_l(\boldsymbol{\nu})\boldsymbol{A}
    \label{ModelMatrix}
\end{equation}
The matrix $M_l$ is a square matrix of size equal to the length of ${z_i}$ that depends on $\boldsymbol{\nu}$. The $i$-th column of $M_l$ contains the elements of the PSF centered on $z_i$ and shifted in frequency by $\nu_i$.

Using the new expression of the model spectrogram in (\ref{ModelMatrix}) in the MAP criterion (\ref{Criterion}) we see the criterion is quadratic with respect to $\boldsymbol{A}$. It is possible to determine the amplitudes that minimize the criterion by canceling its gradient with respect to the amplitudes. The $\boldsymbol{A}$ that minimizes $J$ can be shown to be:
\begin{align}
    \boldsymbol{\hat{A}}(\boldsymbol{\nu})=&\Bigl ( \sum_l (M_l(\boldsymbol{\nu}))^T C_l^{-1} M_l(\boldsymbol{\nu}) \nonumber + \eta\Delta^T\Delta \Bigr )^{-1} \\
    & \Bigl ( \sum_l (M_l(\boldsymbol{\nu}))^T C_l^{-1}S_{\text{exp}}(\cdot,f_l)) \Bigr ) \ ,
\end{align}
where $C_l$ is a diagonal matrix which contains $\sigma^2(\cdot,f_l)$.

Finally, the MAP criterion to be minimized is now $J(\boldsymbol{\hat{A}}(\boldsymbol{\nu}),\boldsymbol{\nu})$ and is simpler because it explicitly depends only on $\boldsymbol{\nu}$.

\section{Inversion results on simulated and experimental spectrograms}

\subsection{Simulated spectrograms with realistic noise}

To accurately simulate the spatial correlation of noise in the spectrogram, we use a random realization of the temporal covariance matrix $Q$ of $\ihet$, computed from the 'feuilleté' model \cite{salamitou1995simulation}. To compute a periodogram, we select a subset $Q_i$ of $Q$ corresponding to the range gate at $z_i$. Then, we derive the covariance matrix of the FT of the lidar signal $R_i=WQ_iW^H$ where $H$ stands for the Hermitian transpose and $W$ is the discrete FT matrix operator. The periodogram for the distance $z_i$ is given by the diagonal elements of $R_i$.

\begin{figure}[h!tb] 
\centering
\includegraphics[width=0.95\linewidth]{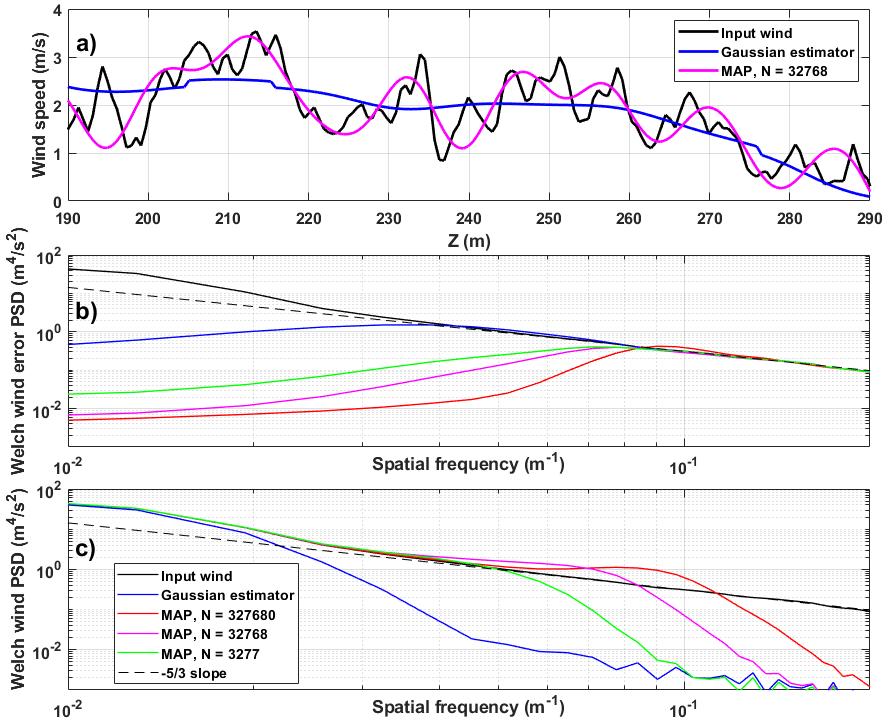}%
  \vspace*{-0.5\baselineskip}%
  \caption{a) Estimation of one wind speed realization by Gaussian estimator and MAP inversion. b) Wind error PSD and input wind speed PSD (black line). c) Reconstructed wind speed PSD from Gaussian estimator and MAP inversion for different average numbers $N$, compared to input wind speed PSD and Kolmogorov law.}
  \label{RWSandPSD}
\end{figure}

The main parameters of the lidar are the following: a sampling frequency of $500\,\text{MHz}$ with periodograms every $0.6\,\text{m}$, a square pulse of width $200\,\text{ns}$, a Gaussian window of $100\,\text{ns}$ full width at half-maximum (FWHM) and a pulse energy of $150\,\text{µJ}$. We apply our method on 30 turbulent wind realizations and we vary the number $N$ of laser shots per wind realization. A variation of $N$ produces a proportional variation of the SNR. The backscattering amplitudes are set constant along the line of sight to have a homogeneous SNR. On each realization a wind speed estimation is made using the Gaussian estimator and is used as the initial vector for the inversion. Then we use our MAP method to obtain a second estimate of the wind speed, and the result for one realization is shown in Figure~\ref{RWSandPSD}a). We see a very significant gain in wind speed estimation for structures of length larger than 15 meters, compared to the Gaussian estimator which only retrieves structures of length larger than 40 meters.

To quantify the spatial resolution (SR) reached we compute the Power Spectral Density (PSD) of the estimated wind speed profile using the Welch spectral estimation method~\cite{Welch}. We define SR as the frequency at which the PSD of the error, i.e. the difference between the estimated wind speed and the input wind speed, reaches 50\% of the input wind PSD. However, in experimental measurements the input wind is unknown. Therefore, we will measure the frequency at which the wind speed PSD drops below 50\% from the -5/3 slope theoretical curve.

The reconstructed PSD and PSD of the error are computed for each realization and averaged to be compared to the average PSD of the input wind speed and to the Kolmogorov law. The PSD of the error is presented in Figure~\ref{RWSandPSD}b) for various values of $N$. The wind field is generated by the inverse FFT of the product of an independent and identically distributed noise with the square root of Von Karman wind PSD \cite{owens1978algorithm}. The overshoot at low frequencies of the input wind speed PSD relative to the -5/3 slope is due to the finite support of the FFT. We see that the error of the wind speed by the MAP is always below the estimation by the Gaussian estimator. The reconstructed PSD is presented in Figure~\ref{RWSandPSD}c) for various values of $N$ and shows a better reconstruction of the wind speed for spatial frequencies larger than $0.03 \, \text{m}^{-1}$.

The estimated SR for the Gaussian estimator is 33 m, which is consistent with the coherent wind lidar definition of the spatial resolution in~\cite{banakh_resolution}, which gives a SR of 36 m with our parameters. At a low SNR with $N=3277$ the estimated SR is 16 m, which is a factor of 2 smaller than the SR of the Gaussian estimator. When $N$ is higher and so is the SNR, the SR is better and is 15 m for a typical averaging number $N=32768$, i.e., a gain of a factor 2.2 in the SR. For $N=327680$, the SR goes down to 13 m. So we have a SR gain of a factor 2.5 at a very high SNR.

In short, our inversion method improves spatial resolution by a factor of 2 at low SNRs and up to 2.5 at high SNRs, as expected, with greater improvement at higher SNRs (larger N). 

\subsection{Experimental spectrograms}

\begin{figure}[h!tb]
\begin{center}\leavevmode
\includegraphics[width=0.85\linewidth]{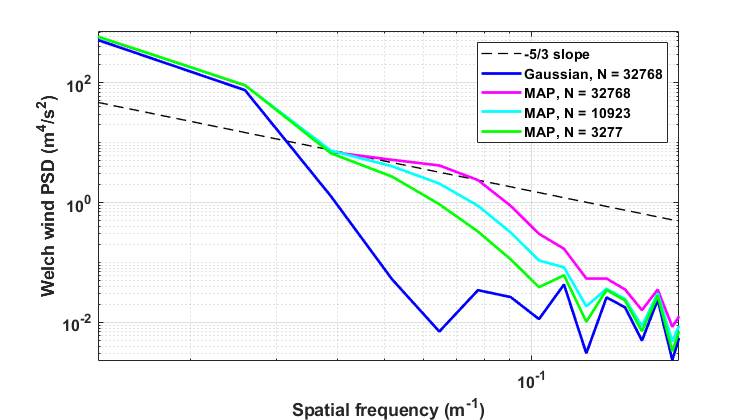}%
  \vspace*{-0.5\baselineskip}%
  \caption{Average PSD on 124 measurements of wind speed estimated by the Gaussian estimator and by the MAP inversion, compared to the Kolmogorov law for different SNRs}
  \label{Kolm_LCP}
  \end{center}
\end{figure}

We apply our algorithm on experimental spectrograms from the ONERA's lidar LCP (\textit{Lidar Courte Portée} for Short Range Lidar). The LCP parameters are the following: a sampling frequency of $500\,\text{MHz}$ with periodograms every $0.3\,\text{m}$, a square pulse of width $200\,\text{ns}$, a Gaussian window of $100\,\text{ns}$ FWHM, a pulse energy of $50\,\text{µJ}$, a telescope diameter of $5\,\text{cm}$, and a pulse repetition frequency of $30\,\text{kHz}$. The measurements are made in the range from 60\,m to 270\,m with a number of laser shots averaged at $N=32768$, and to test the variation of the SNR on the algorithm performance, we also applied it on spectrograms computed on the 10923 and 3277 first laser shots only.
The Gaussian estimator and the MAP are used to estimate wind speed in 124 wind measurements made at ONERA on 17$^{\text{th}}$ September 2024 between 18:07 and 18:13. Figure~\ref{Kolm_LCP} shows the averaged PSD of wind speed for the Gaussian estimator and for the MAP for the three values of $N$. The -5/3 slope of the Kolmogorov law is also plotted to estimate the different SRs, and was checked by computing the temporal PSD of lidar wind speed estimation. 

The estimated SR of the Gaussian estimator is 33 m. This value is consistent with the spatial resolution formula and the value estimated in the simulations. On the other hand, with the MAP inversion we achieve a spatial resolution of 11 m at $N = 32768$ which is a typical averaging number, so a gain in spatial resolution of a factor 3. This improvement is also visible in the reduction of the criterion by a factor 4 (see Supplement Section 2 for more details). The average SNR is 22 dB using the definition of the SNR as the ratio between maximum signal value in one periodogram over the noise standard deviation. In addition, we see that the spatial resolution increases when the SNR is better, since the SR is 15 m for $N=10923$ and 18 m for $N=3277$. The evolution is consistent with the SR estimated in simulations presented in section A. In addition, using Figure~\ref{RWSandPSD}c) and the same method to estimate SR, SR is 16 m at $N=3277$ and 11 m at $N=32768$. SR estimated in experimental measurement is consistent with the values from simulations.

\section{Conclusion}

To our knowledge, this is the first application of a MAP approach to invert spectrogram data from a coherent wind lidar. Our wind speed estimation achieves a gain in spatial resolution of a factor of 2 to 2.5 depending on the SNR. This gain has been confirmed in an experimental measurement using ONERA's lidar LCP by comparison with the Kolmogorov law.

\subsection*{Disclosure}

The authors declare no conflicts of interest.

\subsection*{Data Availability}

Data underlying the results presented in this Letter are
not publicly available at this time but may be obtained from the authors upon reasonable request.

\subsection*{Supplemental document}

See Supplement 
page~\pageref{supplement}
for supporting content.

\bibliography{sample}

\begin{thebibliography}{10}
\newcommand{\enquote}[1]{``#1''}

\bibitem{BeaAirport}
B.~Augere \emph{et~al.}, {\protect\JournalTitle{Measurement Science and Technology Journal}} \textbf{MST-102092.R1} (2015).

\bibitem{TheseMatthieu}
M.~Valla, \enquote{Étude d'un lidar doppler impulsionnel à laser erbium fibré pour des mesures de champ de vent dans la couche limite de l'atmosphère,} Ph.D. thesis, Télécom Paris (2005).

\bibitem{WindEnergy}
P.~Gebraad, J.~Thomas, A.~Ning, \emph{et~al.}, {\protect\JournalTitle{Wind Energy}} \textbf{20} (2016).

\bibitem{AppliMeteo}
M.~Huffaker and M.~Hardesty, \enquote{Remote sensing of atmospheric wind velocities using solid-state and co2 laser systems,} in \emph{Proceedings of {IEEE 84},}  (1996).

\bibitem{smith1972optical}
R.~G. Smith, {\protect\JournalTitle{Applied optics}} \textbf{11}, 2489 (1972).

\bibitem{harris2006lidar}
M.~Harris, M.~Hand, and A.~Wright, \enquote{Lidar for turbine control: March 1, 2005-november 30, 2005,} Tech. rep., National Renewable Energy Lab.(NREL), Golden, CO (United States) (2006).

\bibitem{GaussianFit}
L.~Lombard, M.~Valla, G.~Canat, and A.~Dolfi-Bouteyre, \enquote{Performance of frequency estimators for real time display of high prf pulsed fibered lidar wind map,} in \emph{CLRC 18th Coherent Laser Radar,}  (2016).

\bibitem{MLE}
M.~Valla, B.~Augère, J.-P. Cariou, \emph{et~al.}, \enquote{Fourier transform maximum likelihood estimator for distance resolved velocity measurement with a pulsed 1.55 µm erbium fiber laser based lidar,} in \emph{Proceedings of {$13^{th}$ Coherent Lidar Radar Conference},}  (2005).

\bibitem{EDPST}
C.~Liang, C.~Wang, X.~Xue, \emph{et~al.}, {\protect\JournalTitle{Optics Letters}} \textbf{47}, 3179 (2022).

\bibitem{Wiener2DLidar}
Y.~Zhao, Y.~Zhang, X.~Zhu, \emph{et~al.}, \enquote{A 2d post-processing method for resolution enhancement of coherent wind lidar,} in \emph{Fifteenth International Conference on Information Optics and Photonics (CIOP 2024),} , vol. 13418 (SPIE, 2024), pp. 787--793.

\bibitem{Gretsi}
T.~Martin, M.~Valla, L.~Mugnier, \emph{et~al.}, \enquote{Super-résolution pour les lidars vent hétérodynes par approche inverse,} in \emph{Proceedings of {$29^{th}$ GRETSI},}  (2023).

\bibitem{JTFA}
C.~Wang, H.~Xia, Y.~Liu, \emph{et~al.}, {\protect\JournalTitle{Optics Communications}} \textbf{424}, 48 (2018).

\bibitem{GolayCoding}
C.~Wang, H.~Xia, Y.~Wu, \emph{et~al.}, {\protect\JournalTitle{Optics Letters}} \textbf{44}, 311 (2019).

\bibitem{SubMeterPRCPM}
Y.~Zhang, J.~Yuan, Y.~Wu, \emph{et~al.}, {\protect\JournalTitle{Physical Review Fluids}} \textbf{8}, L022701 (2023).

\bibitem{PSK}
Y.~Zhang, Y.~Wu, and H.~Xia, {\protect\JournalTitle{Journal of Lightwave Technology}} \textbf{40}, 7471 (2022).

\bibitem{salamitou1995simulation}
P.~Salamitou, A.~Dabas, and P.~H. Flamant, {\protect\JournalTitle{Applied optics}} \textbf{34}, 499 (1995).

\bibitem{idier2013bayesian}
J.~Idier, \emph{Bayesian approach to inverse problems} (John Wiley \& Sons, 2013).

\bibitem{Conan-a-98}
J.-M. Conan, L.~M. Mugnier, T.~Fusco, \emph{et~al.}, {\protect\JournalTitle{ao}} \textbf{37}, 4614 (1998).

\bibitem{paul2013}
B.~Paul, L.~Mugnier, J.-F. Sauvage, \emph{et~al.}, {\protect\JournalTitle{Optics Express}} \textbf{21}, 31751 (2013).

\bibitem{jonquiere2025zonal}
H.~Jonqui{\`e}re, L.~M. Mugnier, V.~Michau, and R.~Mercier-Ythier, {\protect\JournalTitle{Optics and Lasers in Engineering}} \textbf{184}, 108615 (2025).

\bibitem{Welch}
P.~Welch, {\protect\JournalTitle{IEEE Transactions on audio and electroacoustics}} \textbf{15}, 70 (1967).

\bibitem{owens1978algorithm}
A.~Owens, {\protect\JournalTitle{Journal of Geophysical Research: Space Physics}} \textbf{83}, 1673 (1978).

\bibitem{banakh_resolution}
V.~Banakh and I.~Smalikho, \emph{Coherent Doppler wind lidars in a turbulent atmosphere} (Artech House, 2013).

\end{thebibliography}

\newpage\label{supplement}


\section*{Supplementary material}

# Bayesian approach for spatial super-resolution of heterodyne wind lidars: supplemental document

## 1. PRIOR EFFECT ON MAP INVERSION

In section "Spectrogram inversion" of the main manuscript we describe a prior on the wind speed based on the Kolmogorov law. Moreover, the prior is not exactly the Kolmogorov law as we supposed a -2 power law for the Power Spectral Density (PSD) and not a -5/3 power law for PSD. This prior aims to avoid an uncontrolled amplification of the noise in the high spatial frequencies. Figure S1 shows a simulation of wind reconstruction at high SNR with and without the use of the prior on the wind speed in order to examplify how the MAP limits noise amplification (spurious peaks), while keeping a notably better resolution than the Gaussian estimator. Note that at low SNR the use of the prior is absolutely necessary.

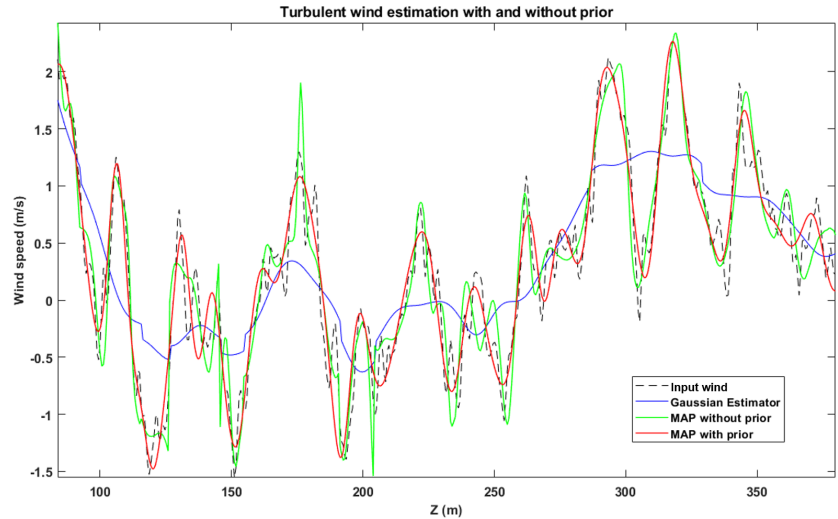

**Fig. S1.** Kolmogorov-type wind speed profile (black dashed line) estimated by Gaussian Estimator (blue line), MAP with (red line) and without (green line) prior on the wind speed

The estimation of wind field without prior has more high spatial frequency glitches than the wind estimated by the Gaussian estimator. But some peaks and little structures are overestimated for instance at 145 m, 180 m and 205 m. Using the prior we avoid these amplified peaks without degrading too much the estimation of high spatial frequencies.

To prove that our prior does not limit the use of our MAP to turbulent wind fields, we show in Figure S2 the reconstruction of an aircraft-induced vortex whose PSD is far from a -5/3 power law.

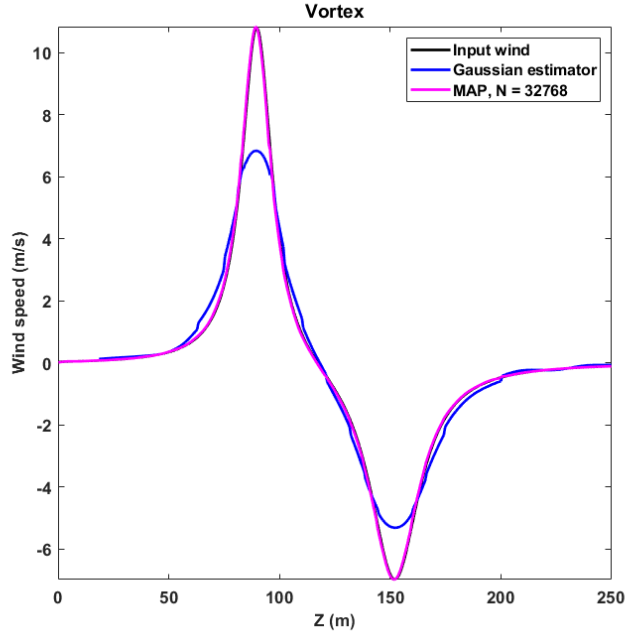

**Fig. S2.** Simulated wind speed estimation at high SNR by Gaussian estimator and MAP for an example of aircraft-induced vortex

Even if the vortex PSD is not the same as we assumed in the prior, the vortex is perfectly estimated by the MAP whereas the Gaussian estimator smooths the input wind.

## 2. VALIDATION OF EXPERIMENTAL DATA BY COMPARING THE DATA SPECTROGRAM TO THE ESTIMATED MODEL

During atmospheric experiments, we do not have access to the true wind profile at the time of measurement. Therefore, for validation we compare the PSD of the estimated wind speed to Kolmogorov law, in order to estimate the spatial resolution gain. Another data we can look into is the difference between the spectrogram model computed from the estimated wind speed and the data spectrogram. We show in Figure S3 a simulation in which the MAP estimated wind values improves the quality of the fit of the model to the data, relative to Gaussian estimated wind values.

The middle right image in Figure S3 shows the error between the model based on Gaussian estimation of wind speed. The error maximum amplitude is 0.15 which is more than an order of magnitude above the noise standard deviation equal to 0.0055. Using the MAP estimation of wind speed, which is closer to the input wind than by the Gaussian estimator as the top left graph in Figure S3 shows, the error amplitude is now below 0.02. It is almost the same amplitude as the noise in the data spectrogram. The MAP criterion has decreased by a factor around 2.6. We show that an improved wind speed estimation induces a decrease of the criterion, and the decrease of the MAP criterion corresponds to an increase of the a posteriori probability of the solution.

We do the same work for experimental data in Figure S4.

As in simulation the wind estimation by the MAP has more little spatial structures than the Gaussian estimator, and the error amplitude of the model spectrogram versus the data spectrogram using the MAP is divided by 10 here in comparison of the error using the Gaussian estimator. The criterion value with the Gaussian estimator is divided by almost 4.5 using the MAP. In addition to the higher cutoff frequency from wind speed PSD relative to the Kolmogorov slope, the improved quality of the fit between the model computed with MAP and spectrogram data reinforces our belief in a spatial resolution gain thanks to the MAP estimator.

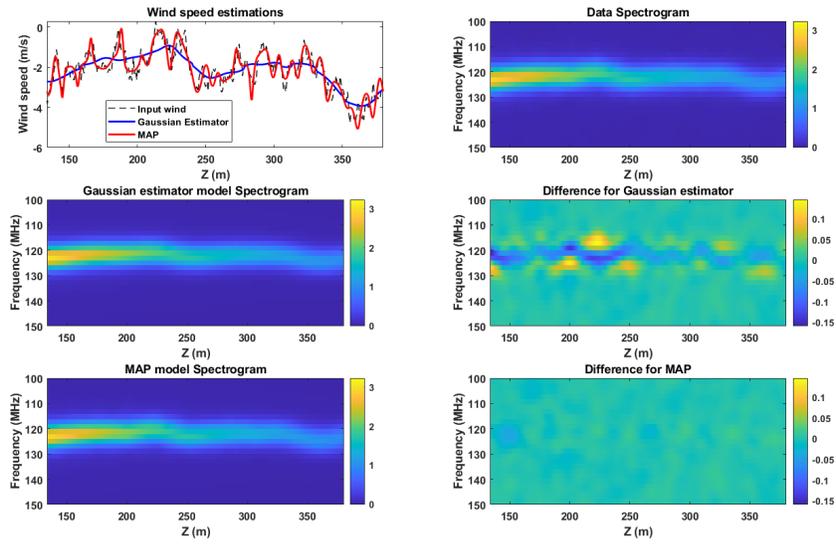

**Fig. S3.** For a simulated spectrogram (top right), wind speed estimations (top left), model spectrogram computed (middle and bottom left), and difference to data (middle and bottom right)

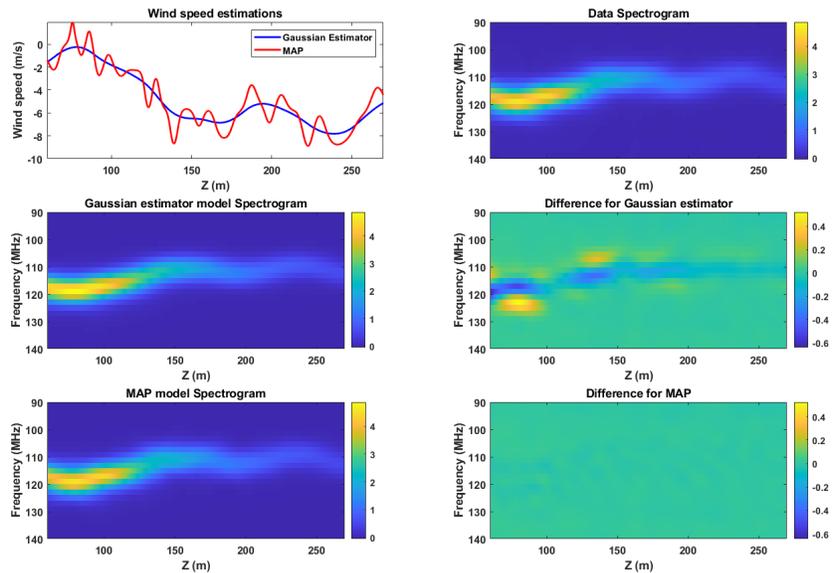

**Fig. S4.** For an experimental spectrogram (top right), wind speed estimations (top left), model spectrogram computed (middle and bottom left), and difference to data (middle and bottom right). Experimental data acquired from a measurement at ONERA Palaiseau with the ONERA's lidar LCP, made on September 13[th] 2024.



\end{document}